\documentclass[showpacs,preprintnumbers,onecolumn,superscriptaddress,aps,nofootinbib]{revtex4}
\usepackage{amssymb}
\usepackage[centertags]{amsmath}
\usepackage{txfonts}
\usepackage{epsfig}
\usepackage{bm}
\usepackage{color}
\usepackage{graphicx,graphics}
\usepackage{multirow}
\usepackage{float}
\usepackage{ulem}
\usepackage[pdfstartview=FitH]{hyperref}
\allowdisplaybreaks[2]   

\begin{document}

\title{Different production sources of light nuclei in relativistic heavy-ion collisions}

\author{Rui-qin Wang}
\affiliation{School of Physics and Engineering, Qufu Normal University, Shandong 273165, China}

\author{Jun Song}
\affiliation{Department of Physics, Jining University, Shandong 273155, China}

\author{Gang Li}
\affiliation{School of Physics and Engineering, Qufu Normal University, Shandong 273165, China}

\author{Feng-lan Shao}
\email {shaofl@mail.sdu.edu.cn}
\affiliation{School of Physics and Engineering, Qufu Normal University, Shandong 273165, China}

\begin{abstract}

We propose a new method, i.e., an exclusive quark combination model + an inclusive hadron recombination model, to study different production sources of light nuclei in relativistic heavy-ion collisions.
We take deuterons and $^3$He produced in Pb-Pb collisions at $\sqrt{s_{NN}}=2.76$ TeV as examples to present contributions of different production sources
by studying their rapidity densities $dN/dy$, yield ratios and transverse momentum ($p_T$) spectra just after the hadronization as well as at the final kinetic freeze-out.
We find that: only very small fractions of $d$ and $^3$He are created just after the hadronization;
nucleons from $\Delta$ resonance decays make a much larger contribution to the regeneration of light nuclei at the hadronic phase stage, and this contribution takes about 79\% and 92\% for $d$ and $^3$He, respectively,
observed at the final kinetic freeze-out. 
We also find that yield ratios $d/p$, $^3$He$/p$ and $^3$He$/d$ are good observables to probe contributions for light nuclei from different production sources, 
and provide a natural explanation for constant values of $d/p$ and $^3$He$/p$ as a function of the averaged charged multiplicity measured by the ALICE Collaboration.
\end{abstract}

\maketitle

\section{Introduction}

The production of light nuclei in relativistic heavy-ion collisions is of importance for many issues in nuclear physics and particle physics.
It not only can help to understand the mechanism of cluster formation in the interior of the fireball of a heavy-ion collision 
but also can serve as an effective probe of the fireball freeze-out properties \cite{Gosset1977PRC,Mekjian1978PRC,Kapusta1980PRC,Nagle1996PRC,LWChen2003NPA}.
Experimental measurements of light nuclei have been extensively executed at the Relativistic Heavy Ion Collider (RHIC) \cite{STARPRL2001,PHENIXPRL2007,STARarXiv2009,STARPRC2016,STARarXiv2017}, 
at relatively low-energy collisions such as those obtained by the NA49 Collaboration at the Super Proton Synchrotron (SPS) \cite{Bearden2002EPJC,NA49PLB2000,NA49PPC2004,NA49PPC2012,NA49PPC2016}, 
and more recently at the very high energy reactions at the Large Hadron Collider (LHC) \cite{deuPRC2016,JPCS2015,ALICE2016NPA,JPCS2017}. 
An interesting phenomenon observed at LHC is that the ratio $d/p$ in Pb-Pb collisions is larger by a factor of about 2.2 than that in pp collisions while
ratios of identified hadrons such as $p/\pi$ and $\Lambda/K_S^0$, etc., do not show significant differences between Pb-Pb and pp collisions \cite{deuPRC2016,FLShao2017PRC}.
Up to now, there is no satisfied understanding for such phenomenon.

Presently, there are two kinds of production mechanisms used to describe light nucleus formation in theory. 
One is the thermal model \cite{Csernai1986PR,AndronicPLB2011,Chatterjee2014PRC} and the other is the recombination/coalescence model \cite{Butler1963PR,Schwarzschild1963PR,SatoPLB1981,Dover1991PRC,YGMa2012PRC,GChen2013PRC,LWChenPLB2015,LLZhu2017PRC}. 
In the recombination/coalescence model, light nuclei can be formed by coalescence of nucleons produced just after the hadronization and/or those from resonance decays. 
Since the binding energies of light nuclei are very small ($\sim$ a few MeV),
final-state coalescence, i.e., nucleons recombining into light nuclei at a
final stage of the hadronic phase evolution (at the final kinetic freeze-out), is commonly adopted
in different recombination/coalescence models
\cite{YGMa2012PRC,GChen2013PRC,LWChenPLB2015,LLZhu2017PRC}. 
In fact, light nuclei can be formed in the whole dynamical process of the hadronic phase evolution. 
At the begining of the hadronic evolution when the temperature is high, light nuclei might be formed and then destroyed immediately due to the violating collisions in the environment. 
As the temperature becomes low, they might be created again by final-state coalescence. 
An important question is whether light nuclei are mainly produced just after the hadronization, i.e., in relatively high temperature environment, or at a later stage of
the hadronic evolution via final-state coalescence \cite{JPCS2016}.

In the present paper, we will make an estimation how many light nuclei can be formed just after the hadronization, 
and to what probability the formed light nuclei can survive until the final kinetic freeze-out. 
Also we will quantitatively see what proportion of the final
observed light nuclei comes from the early formed and surviving ones,
and what proportion of them comes from the regenerated ones in the hadronic evolution stage. 
Here we will employ a hadron recombination
model to make a systematic study of the production of light nuclei just after the hadronization as well as at the final kinetic freeze-out,
based on an exclusive description for identified hadrons with a specific Quark Combination Model developed by ShanDong group SDQCM \cite{QBXie1988PRD,FLShao2005PRC,CEShao2009PRC}.

The rest of the paper is organized as follows. In Sec.~II, we introduce the hadron recombination model dealing with the light nucleus formation.
In Sec.~III, we calculate the rapidity densities $dN/dy$, yield ratios and $p_T$ spectra of $d$ and $^3$He in Pb-Pb collisions at $\sqrt{s_{NN}}=2.76$ TeV just after the hadronization
and those at the final kinetic freeze-out.
We give the results and discussions in this section. Sec.~IV presents our summary.

\section{The hadron recombination model}  \label{model}

After the hadronization of the partonic system produced early in high energy heavy-ion collisions, various kinds of hadrons occur and light nuclei can then be formed through the recombination of protons and neutrons.
The hadron recombination model deals with how nucleons recombine into different light nuclei.
In the hadron recombination model, momentum distributions of $d$ and $^3$He can be expressed as follows
{\setlength\arraycolsep{0.2pt}
\begin{eqnarray}
&& f_{d}(p_d)=  \int dx_1dx_2 dp_1 dp_2  f_{pn}(x_1,x_2;p_1,p_2) \mathcal {R}_{d}(x_1,x_2;p_1,p_2,p_d),      \label{eq:fdgeneral}  \\
&& f_{^3\text{He}}(p_{^3\text{He}})=  \int dx_1dx_2dx_3 dp_1dp_2dp_3  f_{ppn}(x_1,x_2,x_3;p_1,p_2,p_3) \mathcal {R}_{^3\text{He}}(x_1,x_2,x_3;p_1,p_2,p_3,p_{^3\text{He}}).      \label{eq:fHe3general}                        
\end{eqnarray} }%
$f_{pn}(x_1,x_2;p_1,p_2)$ and $f_{ppn}(x_1,x_2,x_3;p_1,p_2,p_3)$ are the two- and three- nucleon joint coordinate-momentum distributions for $(pn)$ and $(ppn)$, respectively. 
Kernel functions $\mathcal {R}_{d}(x_1,x_2;p_1,p_2,p_d)$ and $\mathcal {R}_{^3\text{He}}(x_1,x_2,x_3;p_1,p_2,p_3,p_{^3\text{He}})$ 
denote the probability density for $p$ and $n$ with momenta $p_1$ and $p_2$ at $x_1$ and $x_2$ to recombine into a $d$ of momentum $p_d$
and that for $p$, $p$ and $n$ with momenta $p_1$, $p_2$ and $p_3$ at $x_1$, $x_2$ and $x_3$ to recombine into a $^3$He of momentum $p_{^3\text{He}}$. 
Eqs.~(\ref{eq:fdgeneral}) and (\ref{eq:fHe3general}) are the starting point of describing the production of light nuclei in high energy reactions based on the basic recombination/coalescence idea.

The joint distributions $f_{pn}(x_1,x_2;p_1,p_2)$ and $f_{ppn}(x_1,x_2,x_3;p_1,p_2,p_3)$ are the number densities that satisfy
\begin{eqnarray}
&&\int f_{pn}(x_1,x_2;p_1,p_2) dx_1dx_2 dp_1dp_2 = N_{pn},  \\
&&\int f_{ppn}(x_1,x_2,x_3;p_1,p_2,p_3) dx_1dx_2dx_3 dp_1dp_2dp_3=N_{ppn}, 
\end{eqnarray} 
where $N_{pn}=N_pN_n$ and $N_{ppn}=N_p(N_p-1)N_n \approx N_p^2N_n$
are the numbers of all the possible $(pn)$'s and $(ppn)$'s in the bulk hadronic system that we consider.
$N_p$ and $N_n$ are the number of protons and that of neutrons in the corresponding hadronic system.
We rewrite $f_{pn}(x_1,x_2;p_1,p_2) = N_{pn} f^{(n)}_{pn}(x_1,x_2;p_1,p_2)$ and $f_{ppn}(x_1,x_2,x_3;p_1,p_2,p_3) = N_{ppn} f^{(n)}_{ppn}(x_1,x_2,x_3;p_1,p_2,p_3)$,
so that the coordinate-momentum distributions are normalized to unity where we denote by using the superscript ($n$).
In terms of these normalized joint distributions, we have
{\setlength\arraycolsep{0.2pt}
\begin{eqnarray}
&& f_{d}(p_d)= N_{pn} \int dx_1dx_2 dp_1 dp_2  f^{(n)}_{pn}(x_1,x_2;p_1,p_2) \mathcal {R}_{d}(x_1,x_2;p_1,p_2,p_d),      \label{eq:fdgeneral1}  \\
&& f_{^3\text{He}}(p_{^3\text{He}})= N_{ppn} \int dx_1dx_2dx_3 dp_1dp_2dp_3  f^{(n)}_{ppn}(x_1,x_2,x_3;p_1,p_2,p_3) \mathcal {R}_{^3\text{He}}(x_1,x_2,x_3;p_1,p_2,p_3,p_{^3\text{He}}).  \label{eq:fHe3general1}                        
\end{eqnarray} }%

Kernel functions $\mathcal {R}_{d}(x_1,x_2;p_1,p_2,p_d)$ and $\mathcal {R}_{^3\text{He}}(x_1,x_2,x_3;p_1,p_2,p_3,p_{^3\text{He}})$
carry a lot of kinematical and dynamical information on the nucleons recombining into light nuclei,
but their precise forms are very ambiguous at present due to their complicated non-perturbative properties.
Despite all this, we have known kernel functions should be endowed with the following three points. 
The first is satisfying the momentum conservation, so they should contain the term $\delta(\displaystyle{\sum^2_{i=1}} p_i-p_d)$ or $\delta(\displaystyle{\sum^3_{i=1}} p_i-p_{^3\text{He}})$. 
The second is they should depend on the intrinsic quantum numbers of the formed light nuclei, which are presented by $C_d$ and $C_{^3\text{He}}$.
Considering $A$ nucleons of spin $1/2$ to form a nucleus of angular momentum $J_A$ is $(2J_A+1)/2^A$, we have $C_d=3/4$ and $C_{^3\text{He}}=1/4$. 
The last but the most important is the coordinate and momentum dependencies of these kernel functions.
Even though they can not be solved explicitly from first principles, we know they should increase with the decreases of the relative space positions and the relative momenta of the recombined nucleons.
In general, the coordinate and momentum dependencies of these kernel functions may be coupled to each other. 
But in this paper, we do not study such coupling effects. 
In contrast, in the following, we consider only a simple case where the coordinate and the momentum dependencies of kernel functions are decoupled from each other, i.e., they are factorized.
Based on the above three points, we can write kernel functions as follows
{\setlength\arraycolsep{0.2pt}
\begin{eqnarray}
&&  \mathcal {R}_{d}(x_1,x_2;p_1,p_2,p_d) = C_d \mathcal {R}_{d}^{(x)}(x_1,x_2) \mathcal {R}_{d}^{(p)}(p_1,p_2) \delta(\displaystyle{\sum^2_{i=1}} p_i-p_d),      \label{eq:Rdfac}  \\
&&  \mathcal {R}_{^3\text{He}}(x_1,x_2,x_3;p_1,p_2,p_3,p_{^3\text{He}}) = C_{^3\text{He}} \mathcal {R}_{^3\text{He}}^{(x)}(x_1,x_2,x_3) 
                      \mathcal {R}_{^3\text{He}}^{(p)}(p_1,p_2,p_3)  \delta(\displaystyle{\sum^3_{i=1}} p_i-p_{^3\text{He}}).      \label{eq:RHe3fac}                        
\end{eqnarray} }%
The coordinate and momentum dependence parts are adopted as experiential Gaussian forms $\mathcal {R}^{(x)}_{d}(x_1,x_2)=e^{-\frac{(x_1-x_2)^2}{2R_d^2}}$,
$\mathcal {R}^{(x)}_{^3\text{He}}(x_1,x_2,x_3)=e^{-\frac{(x_1-x_2)^2}{2R_{^3\text{He}}^2}}  e^{-\frac{(x_1+x_2-2x_3)^2}{6R_{^3\text{He}}^2}}$
and $\mathcal {R}^{(p)}_{d}(p_{1},p_{2})=e^{-\frac{(p_{1}-p_{2})^2}{2\sigma_d^2}}$,
$\mathcal {R}^{(p)}_{^3\text{He}}(p_1,p_2,p_3)=e^{-\frac{(p_1-p_2)^2}{2\sigma_{^3\text{He}}^2}}  e^{-\frac{(p_1+p_2-2p_3)^2}{6\sigma_{^3\text{He}}^2}}$, 
where the widths of the Gaussian distributions are the root-mean-square radii $R_d=1.96$ fm, $R_{^3\text{He}}=1.76$ fm \cite{radii2009PRC}
and $\sigma_d=1/\sqrt{\frac{4}{3} R_d^2}=0.087$ GeV, $\sigma_{^3\text{He}}=1/\sqrt{R_{^3\text{He}}^2}=0.112$ GeV \cite{LWChenPLB2015}.

We further assume that the normalized joint distributions of the nucleons are coordinate and momentum factorized as follows
{\setlength\arraycolsep{1pt}
\begin{eqnarray}
 &&f^{(n)}_{pn}(x_1,x_2;p_1,p_2) = f^{(n)}_{pn}(x_1,x_2)   f^{(n)}_{pn}(p_1,p_2),  \label{eq:fpnfac} \\   
 &&f^{(n)}_{ppn}(x_1,x_2,x_3;p_1,p_2,p_3) = f^{(n)}_{ppn}(x_1,x_2,x_3)   f^{(n)}_{ppn}(p_1,p_2,p_3).   \label{eq:fppnfac}
\end{eqnarray}}%
Substituting Eqs.~(\ref{eq:Rdfac}-\ref{eq:fppnfac}) into Eqs.~(\ref{eq:fdgeneral1}) and (\ref{eq:fHe3general1}), we have 
{\setlength\arraycolsep{0.2pt}
\begin{eqnarray}
&& f_{d}(p_d)= N_{pn} C_{d} \int dx_1dx_2 f^{(n)}_{pn}(x_1,x_2) \mathcal {R}^{(x)}_{d}(x_1,x_2)
                      \int dp_1dp_2 f^{(n)}_{pn}(p_1,p_2) \mathcal {R}^{(p)}_{d}(p_1,p_2) \delta(\displaystyle{\sum^2_{i=1}} p_i-p_d),      \label{eq:fd}  \\
&& f_{^3\text{He}}(p_{^3\text{He}})= N_{ppn} C_{^3\text{He}} \int dx_1dx_2dx_3 f^{(n)}_{ppn}(x_1,x_2,x_3) \mathcal {R}^{(x)}_{^3\text{He}}(x_1,x_2,x_3)  \nonumber  \\
&& ~~~~~~~~~~~~~~~~~~~~~~~~~  \times\int dp_1dp_2dp_3 f^{(n)}_{ppn}(p_1,p_2,p_3) \mathcal {R}^{(p)}_{^3\text{He}}(p_1,p_2,p_3) \delta(\displaystyle{\sum^3_{i=1}} p_i-p_{^3\text{He}}).  \label{eq:fHe3}                        
\end{eqnarray} }%
We denote the coordinate integrals in Eqs.~(\ref{eq:fd}) and (\ref{eq:fHe3}) to be $\mathcal {A}_d$ and $\mathcal {A}_{^3\text{He}}$, respectively, to get
{\setlength\arraycolsep{0.2pt}
\begin{eqnarray}
&& f_{d}(p_d)= N_{pn} C_{d} \mathcal {A}_d
                      \int dp_1dp_2 f^{(n)}_{pn}(p_1,p_2) \mathcal {R}^{(p)}_{d}(p_1,p_2) \delta(\displaystyle{\sum^2_{i=1}} p_i-p_d),      \label{eq:fdA}  \\
&& f_{^3\text{He}}(p_{^3\text{He}})= N_{ppn} C_{^3\text{He}} \mathcal {A}_{^3\text{He}}
             \int dp_1dp_2dp_3 f^{(n)}_{ppn}(p_1,p_2,p_3) \mathcal {R}^{(p)}_{^3\text{He}}(p_1,p_2,p_3) \delta(\displaystyle{\sum^3_{i=1}} p_i-p_{^3\text{He}}),  \label{eq:fHe3A}                        
\end{eqnarray} }%
where
{\setlength\arraycolsep{0.2pt}
\begin{eqnarray}
&& \mathcal {A}_d =  \int dx_1dx_2 f^{(n)}_{pn}(x_1,x_2) e^{-\frac{(x_1-x_2)^2}{2R_d^2}},      \label{eq:Ad}   \\
&& \mathcal {A}_{^3\text{He}} =  \int dx_1dx_2dx_3 f^{(n)}_{ppn}(x_1,x_2,x_3) e^{-\frac{(x_1-x_2)^2}{2R_{^3\text{He}}^2}}  e^{-\frac{(x_1+x_2-2x_3)^2}{6R_{^3\text{He}}^2}}.      \label{eq:AHe3}  
\end{eqnarray} }%
We change the integral variables in Eqs.~(\ref{eq:Ad}) and (\ref{eq:AHe3}) to be
{\setlength\arraycolsep{0.2pt}
\begin{eqnarray}
&& r_1= \frac{x_1-x_2}{\sqrt{2}},       \\
&& r_2= \frac{x_1+x_2}{\sqrt{2}},      
\end{eqnarray} }%
and
{\setlength\arraycolsep{0.2pt}
\begin{eqnarray}
&& r_1= \frac{x_1-x_2}{\sqrt{2}},         \\
&& r_2= \frac{x_1-x_3}{\sqrt{2}},        \\
&& r_3= \frac{x_1+x_2+x_3}{\sqrt{3}},      
\end{eqnarray} }%
respectively, and we can obtain
{\setlength\arraycolsep{0.2pt}
\begin{eqnarray}
&& \mathcal {A}_d =  \int dr_1dr_2 f_{pn}(r_1,r_2) e^{-\frac{r_1^2}{R_d^2}},   \label{eq:Adr}   \\
&& \mathcal {A}_{^3\text{He}} = \frac{2}{\sqrt{3}} \int dr_1dr_2dr_3 f_{ppn}(r_1,r_2,r_3) e^{-\frac{r_1^2}{R_{^3\text{He}}^2}}  e^{-\frac{(2r_2-r_1)^2}{3R_{^3\text{He}}^2}},  \label{eq:AHe3r}  
\end{eqnarray} }%
and the normalization constraints 
{\setlength\arraycolsep{0.2pt}
\begin{eqnarray}
&&   \int f_{pn}(r_1,r_2)  dr_1dr_2=1,   \\
&&   \int \frac{2}{\sqrt{3}} f_{ppn}(r_1,r_2,r_3)  dr_1dr_2dr_3=1.  
\end{eqnarray} }%
To further evaluate $\mathcal {A}_d$ and $\mathcal {A}_{^3\text{He}}$, we assume different coordinate variables are decoupled from each other in the coordinate distributions and have
{\setlength\arraycolsep{0.2pt}
\begin{eqnarray}
&& f_{pn}(r_1,r_2) = f^{(n)}_{pn}(r_1) f^{(n)}_{pn}(r_2),    \label{eq:fpnr}  \\
&& \frac{2}{\sqrt{3}}f_{ppn}(r_1,r_2,r_3) = f^{(n)}_{ppn}(r_1) f^{(n)}_{ppn}(r_2) f^{(n)}_{ppn}(r_3).  \label{eq:fppnr}
\end{eqnarray} }%
The normalized distribution of the relative space positions of the nucleons are
$f^{(n)}_{pn}(r_1) = \frac{1}{(2\pi R_f^2)^{0.5}} e^{-\frac{r_1^2}{2R_f^2}}$, $f^{(n)}_{ppn}(r_1) = \frac{1}{(2\pi R_f^2)^{0.5}} e^{-\frac{r_1^2}{2R_f^2}}$ 
and $f^{(n)}_{ppn}(r_2) = \frac{1}{(2\pi R_f^2)^{0.5}} e^{-\frac{r_2^2}{2R_f^2}}$ \cite{Rche2017acta},
where $R_f$ is the fireball radius at freeze-out.
Substituting Eqs.~(\ref{eq:fpnr}) and (\ref{eq:fppnr}) into Eqs.~(\ref{eq:Adr}) and (\ref{eq:AHe3r}), we have
{\setlength\arraycolsep{0.2pt}
\begin{eqnarray}
&& \mathcal {A}_d =  \frac{R_d}{\sqrt{2}R_f} (1+\frac{R_d^2}{2R_f^2})^{-1/2},   \label{eq:Adfin}   \\
&& \mathcal {A}_{^3\text{He}} = \frac{\sqrt{3}R_{^3\text{He}}^2}{4R_f^2} (1+\frac{3R_{^3\text{He}}^2}{12R_f^2})^{-1/2}  (1+\frac{3R_{^3\text{He}}^2}{4R_f^2})^{-1/2} .  \label{eq:AHe3fin}  
\end{eqnarray} }%
$R_f$ is about 6.1 fm just after the hadronization and 7.0 fm at the final kinetic freeze-out \cite{Rche2017acta,Rkin2016PRC}. 
With these values, the evaluated results for $\mathcal {A}_d$ are 0.2216 just after the hadronization and 0.1942 at the final kinetic freeze-out, and those for
$\mathcal {A}_{^3\text{He}}$ are 0.0346 and 0.0265, respectively.

So far with Eqs.~(\ref{eq:fdA}) and (\ref{eq:fHe3A}), the production of light nuclei just after the hadronization as well as at the final kinetic freeze-out can be computed 
as the corresponding joint-momentum distributions $f_{pn}(p_1,p_2)$ and $f_{ppn}(p_1,p_2,p_3)$ are known.
Ignoring the correlations of nucleons, we have $f_{pn}(p_1,p_2)=f_{p}(p_1)f_{n}(p_2)$ and
$f_{ppn}(p_1,p_2,p_3)=f_{p}(p_1)f_{p}(p_2)f_{n}(p_3)$, where the single nucleon momentum distributions $f_p$ and $f_n$ can be easily computed by the SDQCM.
SDQCM is unique for its ability to exclusively explain yields, rapidity distributions, and $p_T$ spectra of different hadrons \cite{FLShao2005PRC,CEShao2009PRC},
and has many successful applications in $e^+e^-$ annihilations, pp reactions and heavy-ion collisions \cite{QBXie1988PRD,FLShao2005PRC,CEShao2009PRC,QWang2000PRD,RQWang2012PRC,FLShao2017PRC,JSong2017PRC}.
For more detailed information about SDQCM, one can refer to Refs. \cite{FLShao2005PRC,CEShao2009PRC}.

As a brief summary of this section, we want to state that with the exclusive quark combination model SDQCM to obtain momentum distributions of different light, strange and heavy flavor hadrons,
one can apply the hadron recombination model to investigating the production of light nuclei, hypernuclei and other molecular states. 
There are no extra free parameters in this method besides several assumptions and/or approximations explicitly stated above
such as the coordinate and the momentum factorization for the kernel function and the joint distribution function, and the decoupling of different coordinate variables in the coordinate distributions.
Even though it is very simple, this method, an exclusive quark combination model + an inclusive hadron recombination model, provides an effective way to study different production sources of light nuclei,
hypernuclei and other molecular states.

\section{Results and discussions}

In this section, we systematically study the production of light nuclei at the midrapidity in central Pb-Pb collisions at $\sqrt{s_{NN}}= 2.76$ TeV.
Firstly, we need the momentum distributions of constituent quarks just before the hadronization to compute the momentum distributions of identified hadrons with the SDQCM.
Due to focusing on the midrapidity region at so high LHC energy, influences of net quarks from the incident colliding nuclei are ignored.
Also, the isospin symmetry for $u$ and $d$ quarks is adopted.
The input values for the rapidity densities of $u$, $d$ and $s$ quarks are $dN_u/dy=dN_d/dy=679$ and $dN_s/dy=272$, respectively,
and the longitudinal rapidity distributions for both $u$, $d$ and $s$ quarks are
choosed to be uniform since we focus on the hadron production at the rapidity window $|y|<0.5$.
For the transverse momentum distributions of light and strange quarks, we use an exponentially parameterized pattern
$dN/(p_Tdp_T) \propto \exp(-\sqrt{p_T^2+m^2}/T)$ because of focusing on the thermal hadron production. 
The slope parameter $T$ for $u$, $d$ quarks is 0.37 GeV and that for $s$ quarks is 0.40 GeV. 
These values are slightly higher to better describe proton production than those have been obtained in our previous work 
where the production of various light, strange and charm hadrons were studied \cite{RQWang2015PRC}.

With the above inputs, rapidity densities $dN/dy$ and $p_T$ distributions for directly produced hadrons, i.e., those hadrons produced just after the hadronization can be calculated by the SDQCM.
Only including very short lifetime $\rho$, $K^*$, $\Delta$ and $\Sigma^*$ decays, we obtain $dN/dy$ and $p_T$ distributions of hadrons at the final kinetic freeze-out.
We want to state a point about the process begin from just after the hadronization to the final kinetic freeze-out. 
During this process strong decays for some harons such as $\Xi^*$, electromagnetic decays and weak decays are not included because their lifetimes are relatively long so they are set to decay after the kinetic freeze-out.
After considering all strong and electromagnetic decays, we have $dN/dy$ and $p_T$ distributions for the total final state hadrons which can be compared to the LHC experimental measurements.

Rapidity densities $dN/dy$ of identified hadrons in central Pb-Pb collisions at $\sqrt{s_{NN}}=2.76$ TeV are listed in Table \ref{tab:Nhadron}. 
The data in the second column are from Refs. \cite{pikpPRL2012,Ks0lam2013,xiome2013,phipt}.
The third column presents the results for directly produced hadrons.
The fourth column gives the results for hadrons at the final kinetic freeze-out.
The last column is the results for total final state ones which agree with the experimental data and are comparable to our previous results calculated by analytical formulas \cite{RQWang2015PRC}.
Results for $\phi$ and $\Omega$ of directly produced, created at final kinetic freeze-out and total final state ones are the same, since there are no other decay contributions to them.
The directly produced $K/\pi$ ratio is about 0.4, which reveals the strangeness in heavy-ion collisions and is comparable to our previous studies \cite{CEShao2009PRC}.
Directly produced protons take about half of the total final states ones, and the other half comes from $\Delta$ decays.
The lifetime of $\Delta$ is so short that they decay before the final kinetic freeze-out occurs. This leads to the result of protons at kinetic freeze-out equals to that of total final state ones.
That is the reason why some coalescence models can employ the momentum distribution of protons measured experimentally as input when studying the production of light nuclei \cite{LWChenPLB2015}.
$\Lambda$ baryons come from the directly produced or $\Sigma^*$ and $\Sigma^0$ decays. So total final state ones are more than kinetic freeze-out ones, and both are more than directly produced ones.
Only $\Xi^*$ can decay to $\Xi$, and the decay takes place after the kinetic freeze-out. So the directly produced $\Xi$ and those at the kinetic freeze-out are the same, and both of them are less than total final state ones.
For more detailed information of hadron decays, one can refer to Ref. \cite{PDG2016}.

\begin{table*}[htbp]
\renewcommand{\arraystretch}{1.5}
\caption{Rapidity densities $dN/dy$ of identified hadrons in central Pb-Pb collisions at
$\sqrt{s_{NN}}=2.76$ TeV. The data are from Refs. \cite{pikpPRL2012,Ks0lam2013,xiome2013,phipt}.
The last three columns are the results calculated by SDQCM for directly produced hadrons, hadrons freezed out at kinetic freeze-out and total final state ones, respectively.}
\begin{tabular}{p{50pt}p{100pt}p{80pt}p{80pt}p{60pt}}
\toprule
Hadron         &Data                 &SDQCM (DP)       &SDQCM (KI)    &SDQCM (TFS) \\
\colrule
$\pi^+$        &$733\pm54$           &55               &501           &737        \\
$\pi^-$        &$732\pm52$           &55               &501           &737        \\
$K^+$          &$109\pm9$            &22               &88            &111       \\
$K^-$          &$109\pm9$            &22               &88            &111        \\
$K_S^{0}$      &$110\pm10$           &0                &0             &106       \\
$\phi$         &$13.8\pm0.5\pm1.7$   &13.2             &13.2          &13.2     \\
$p$            &$34\pm3$             &16               &33            &33      \\
$\bar p$       &$33\pm3$             &16               &33            &33      \\
$\Lambda$      &$26\pm3$             &7.8              &17            &25    \\
$\bar\Lambda$  &$---$                &7.8              &17            &25    \\
$\Xi^-$        &$3.34\pm0.06\pm0.24$ &2.60             &2.60          &3.90  \\
$\bar\Xi^+$    &$3.28\pm0.06\pm0.23$ &2.60             &2.60          &3.90  \\
$\Omega^-$     &$0.58\pm0.04\pm0.09$ &0.52             &0.52          &0.52     \\
$\bar\Omega^+$ &$0.60\pm0.05\pm0.09$ &0.52             &0.52          &0.52     \\
\botrule
\end{tabular} \label{tab:Nhadron}
\end{table*}

\begin{figure}[htbp]
\centering
 \includegraphics[width=0.8\linewidth]{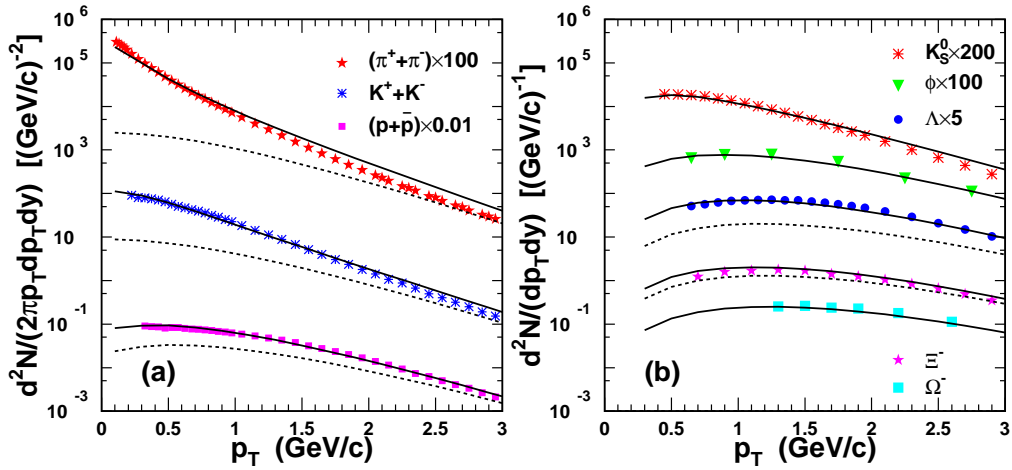}\\
 \caption{(Color online) $p_T$ distributions of identified hadrons at midrapidity in central Pb-Pb collisions at $\sqrt{s_{NN}}=2.76$ TeV.
 The solid lines are the results for total final state hadrons, and the dashed lines are for directly produced ones.
 The filled symbols are the experimental data from Refs. \cite{pikpPRL2012,Ks0lam2013,xiome2013,phipt}.}
 \label{centralHPT}
\end{figure}

$p_T$ distributions of identified hadrons are also computed and the results are in Fig.~\ref{centralHPT}. 
The filled symbols are the experimental data from Refs. \cite{pikpPRL2012,Ks0lam2013,xiome2013,phipt}.
Solid lines are the results for total final state hadrons, and dashed lines are for directly produced ones.
Results for total final state baryons $p$, $\bar p$, $\Lambda$, $\Xi^-$, $\Omega^-$ and $\phi$ mesons agree with the experiential data very well.
Results for total final state $K^\pm$ and $K_S^0$ mesons are slightly harder at $p_T>2$ GeV, and for $\pi^\pm$ the deviation is more serious.
This is due to the decay contributions from other hadrons at relatively large $p_T$ area, where only the exponential quark distribution is not enough. 
Such deviations can be eliminated when include mini-jet quark contribution at relatively large $p_T$ area, and the corresponding work has been done in another paper \cite{RQWang2015PRC}.
In the present paper, we focus on hadrons produced at $p_T<2$ GeV, so the simple exponential quark distribution is enough.
The perfect results for directly produced $p$ and total final state ones (same as kinetic freeze-out ones) are only necessary for the next study of light nuclei.

Based on the good performance of SDQCM for identified hadrons, especially for protons, we turn to the study of light nuclei. 
We calculate rapidity densities $dN/dy$ of $d$ and $^3$He just after the hadronization as well as at the final kinetic freeze-out.
The input nucleon momentum distributions just after the hadronization are $f_{pn}(p_1,p_2)=f_p^{(dir)}(p_1)f_n^{(dir)}(p_2)$ and 
$f_{ppn}(p_1,p_2,p_3)=f_p^{(dir)}(p_1)f_p^{(dir)}(p_2)f_n^{(dir)}(p_3)$, where the superscript $(dir)$ denotes directly produced nucleons.
The input momentum distributions at the final kinetic freeze-out are as follows
{\setlength\arraycolsep{0.2pt}
\begin{eqnarray}
&& f_{pn}(p_1,p_2)=f_p^{(kin)}(p_1)f_n^{(kin)}(p_2),   \\
&& f_{ppn}(p_1,p_2,p_3)=f_p^{(kin)}(p_1)f_p^{(kin)}(p_2)f_n^{(kin)}(p_3). 
\end{eqnarray} }%
The superscript $(kin)$ denotes nucleons freezed out at the final kinetic freeze-out, and they are made of two parts. 
One is the directly produced ones, and the other is from $\Delta$ decays denoted by the superscript $(dec)$, i.e.,
{\setlength\arraycolsep{0.2pt}
\begin{eqnarray}
&& f_{pn}(p_1,p_2)=   [f_p^{(dir)}(p_1)+f_p^{(dec)}(p_1)]    [f_n^{(dir)}(p_2)+f_n^{(dec)}(p_2)]  \nonumber  \\
&&~~~~~~~~~~~~~~~~~ = f_p^{(dir)}(p_1)f_n^{(dir)}(p_2)   +f_p^{(dir)}(p_1)f_n^{(dec)}(p_2)   +f_p^{(dec)}(p_1)f_n^{(dir)}(p_2)    +f_p^{(dec)}(p_1) f_n^{(dec)}(p_2),    \label{eq:fpn}                      \\
&& f_{ppn}(p_1,p_2,p_3)=[f_p^{(dir)}(p_1)+f_p^{(dec)}(p_1)]   [f_p^{(dir)}(p_2)+f_p^{(dec)}(p_2)]   [f_n^{(dir)}(p_3)+f_n^{(dec)}(p_3)]  \nonumber   \\
&&~~~~~~~~~~~~~~~~~~~~~~~~ = f_p^{(dir)}(p_1)f_p^{(dir)}(p_2)f_n^{(dir)}(p_3)  +f_p^{(dir)}(p_1)f_p^{(dir)}(p_2)f_n^{(dec)}(p_3)  +f_p^{(dir)}(p_1)f_p^{(dec)}(p_2)f_n^{(dir)}(p_3)   \nonumber   \\
&&~~~~~~~~~~~~~~~~~~~~~~~~  +f_p^{(dec)}(p_1)f_p^{(dir)}(p_2)f_n^{(dir)}(p_3)+f_p^{(dir)}(p_1)f_p^{(dec)}(p_2)f_n^{(dec)}(p_3)  +f_p^{(dec)}(p_1)f_p^{(dir)}(p_2)f_n^{(dec)}(p_3)  \nonumber \\
&&~~~~~~~~~~~~~~~~~~~~~~~~  +f_p^{(dec)}(p_1)f_p^{(dec)}(p_2)f_n^{(dir)}(p_3)  +f_p^{(dec)}(p_1)f_p^{(dec)}(p_2)f_n^{(dec)}(p_3).   \label{eq:fppn}
\end{eqnarray} }%
The first terms in Eqs.~(\ref{eq:fpn}) and (\ref{eq:fppn}) contribute to light nuclei defined as surviving ones which used to be produced just after the hadronization, 
and all the other terms in Eqs.~(\ref{eq:fpn}) and (\ref{eq:fppn}) contribute to light nuclei defined as regenerated ones.

So far with the momentum distributions of directly-produced protons and those at final kinetic freeze-out (i.e., total final state ones) shown in Fig.~\ref{centralHPT},
we can compute $dN/dy$ and $p_T$ spectra of $d$ and $^3$He.
Momentum distributions of neutrons are the same as those of protons due to adopting isospin symmetry.
The third and fourth columns in Table \ref{tab:Ndeuteron} are the results for just after the hadronization and at the final kinetic freeze-out, respectively.
The experimental data in the second column are from Ref. \cite{deuPRC2016}. 
From Table \ref{tab:Ndeuteron}, one can see that the results computed just after the hadronization are much lower than the data and
the results at the final kinetic freeze-out agree well with the data.
This shows that most of $d$ and $^3$He are produced at the later hadronic stage instead of just after the hadronization.
The number of the surviving $d$ and $^3$He used to be formed just after the hadronization and 
those regenerated ones at the final kinetic freeze-out are also counted, and the corresponding results are in the last two columns.
For $d$ produced just after the hadronization, about 88\% are surviving during the hadron gas evolution, but for $^3$He about 76\% are surviving. 
This is because the formed nuclei are usually destroyed immediately after their formation due to the violating collisions in the high temperature environment,
and it is much easier for two decomposed nucleons recombining into $d$ than three decomposed nucleons recombining into $^3$He when the temperature becomes low.
For the total light nuclei at the final kinetic freeze-out, about 79\% of $d$ are regenerated and 92\% of $^3$He are regenerated. 
These numerical results explicitly show different production sources of $d$ and $^3$He. 

\begin{table*}[htbp]
\renewcommand{\arraystretch}{1.5}
\caption{Rapidity densities $dN/dy$ of $d$ and $^3$He in central Pb-Pb collisions at $\sqrt{s_{NN}}=2.76$ TeV. 
The data are from Ref. \cite{deuPRC2016}.
The third and fourth columns are the results for just after the hadronization and at the final kinetic freeze-out, respectively.
The surviving light nuclei and regenerated ones at final kinetic freeze-out are in the last two columns, respectively.}
\begin{tabular}{p{20pt}p{110pt}p{100pt}p{60pt}p{85pt}p{83pt}}
\toprule
         &Data                                  &Just after Hadronization              &Kinetic                &Surviving at Kinetic          &Regenerated at Kinetic     \\
\colrule
$d$      &$(9.82\pm0.04\pm1.58)\times 10^{-2}$  &$2.59\times 10^{-2}$  &$10.83\times 10^{-2}$  &$2.27\times 10^{-2}$         &$8.56\times 10^{-2}$        \\
$^3$He   &$(2.76\pm0.09\pm0.62)\times 10^{-4}$  &$0.25\times 10^{-4}$  &$2.30\times 10^{-4}$   &$0.19\times 10^{-4}$         &$2.11\times 10^{-4}$         \\
\botrule
\end{tabular} \label{tab:Ndeuteron}
\end{table*}

A series of interesting observables are yield ratios about light nuclei, such as $d/p$, $^3$He$/p$ and $^3$He$/d$.
Recalling Eqs.~(\ref{eq:fdA}) and (\ref{eq:fHe3A}), we have
{\setlength\arraycolsep{0.2pt}
\begin{eqnarray}
&& d/p = N_{p} C_{d} \mathcal {A}_d
                      \int dp_1dp_2 f^{(n)}_{pn}(p_1,p_2) \mathcal {R}^{(p)}_{d}(p_1,p_2),      \label{eq:Rdp}  \\
&& ^3\text{He}/p = N_{p}^2 C_{^3\text{He}} \mathcal {A}_{^3\text{He}}
             \int dp_1dp_2dp_3 f^{(n)}_{ppn}(p_1,p_2,p_3) \mathcal {R}^{(p)}_{^3\text{He}}(p_1,p_2,p_3) ,  \label{eq:RHe3p}   \\   
&& ^3\text{He}/d = N_{p} \frac{C_{^3\text{He}}}{C_d}  \frac{\mathcal {A}_{^3\text{He}}}{\mathcal {A}_d}
             \frac{\int dp_1dp_2dp_3 f^{(n)}_{ppn}(p_1,p_2,p_3) \mathcal {R}^{(p)}_{^3\text{He}}(p_1,p_2,p_3)}
                  {\int dp_1dp_2 f^{(n)}_{pn}(p_1,p_2) \mathcal {R}^{(p)}_{d}(p_1,p_2)}.  \label{eq:RHe3d}             
\end{eqnarray} }%
From Eqs.~(\ref{eq:Adfin}) and (\ref{eq:AHe3fin}), one can get $\mathcal {A}_d  \approx \frac{R_d}{\sqrt{2}R_f}$ 
and $\mathcal {A}_{^3\text{He}} \approx \frac{\sqrt{3}R_{^3\text{He}}^2}{4R_f^2}$ because $({R_d}/{R_f})^2<<1$ and $({R_{^3\text{He}}}/{R_f})^2<<1$.
Momentum integrals in these three ratios are very weakly dependent of the collision centrality considering that the normalized nucleon momentum distributions do not show significant differences in
different centrality collisions.
$C_{d}$ and $C_{^3\text{He}}$ only depend on the intrinsic quantum numbers of light nuclei and are independent of the collision energy and collision centrality.
Therefore, from Eqs.~(\ref{eq:Rdp}-\ref{eq:RHe3d}), one can see that these three ratios are approximately proportional to nucleon densities instead of nucleon numbers, 
and this can well explain the constant values for $d/p$ and $^3$He$/p$ as a function of the averaged charged multiplicity observed by the ALICE Collaboration \cite{deuPRC2016}.
Numerical values of $d/p$, $^3$He$/p$ and $^3$He$/d$ for just after the hadronization and at the final kinetic freeze-out are calculated, respectively, 
and results are in the third and fourth columns in Table \ref{tab:ratio}. Experimental data in the second column in Table \ref{tab:ratio} are from Ref. \cite{deuPRC2016}.

\begin{table*}[htbp]
\renewcommand{\arraystretch}{1.5}
\caption{Yield ratios $d/p$, $^3$He$/p$ and $^3$He$/d$ in central Pb-Pb collisions at $\sqrt{s_{NN}}=2.76$ TeV. 
The data in the second column are from Ref. \cite{deuPRC2016}.
The third and fourth columns are the results for just after the hadronization and at the final kinetic freeze-out, respectively.}
\begin{tabular}{p{40pt}p{145pt}p{110pt}p{65pt}}
\toprule
             &Data                                        &Just after the Hadronization        &Kinetic                     \\
\colrule
$d/p$        &$(3.2188\pm0.0142\pm0.6067)\times 10^{-3}$  &$1.62\times 10^{-3}$            &$3.28\times 10^{-3}$          \\
$^3$He$/p$   &$(1.0611\pm0.0359\pm0.2570)\times 10^{-5}$  &$0.16\times 10^{-5}$            &$0.70\times 10^{-5}$     \\
$^3$He$/d$   &$(2.81\pm0.86)\times 10^{-3}$               &$0.97\times 10^{-3}$            &$2.12\times 10^{-3}$      \\
\botrule
\end{tabular} \label{tab:ratio}
\end{table*}

From expressions for $\mathcal {A}$, we also know values of $\mathcal {A}$ just after the hadronization are slightly larger than those at the final kinetic freeze-out
because of larger fireball radius at the final kinetic freeze-out. Momentum integral parts in Eqs.~(\ref{eq:Rdp}) and (\ref{eq:RHe3p}) are slightly lower just after the hadronization
than those at the final kinetic freeze-out since momentum distributions of directly produced nucleons are harder than those including resonance decay contributions at the final kinetic freeze-out.
Therefore, the ratio of these ratio values for just after the hadronization to their values at the final kinetic freeze-out are mainly determined by the number of protons for just after the hadronization and
that at the final kinetic freeze-out.
From Table \ref{tab:Nhadron}, we have $N_p^{(DP)}/N_p^{(KI)}\approx 1/2$, so
{\setlength\arraycolsep{0.2pt}
\begin{eqnarray}
&& \frac { (d/p)_{JAH} }  {(d/p)_{KI}} \approx \frac{N_p^{(DP)}}{N_p^{(KI)}} \approx \frac{1}{2},   \label{eq:RdpCK} \\
&& \frac { (^3\text{He}/p)_{JAH} }   {(^3\text{He}/p)_{KI}}  \approx  (\frac{N_p^{(DP)}}{N_p^{(KI)}})^2   \approx \frac{1}{4},  \label{eq:RHe3pCK} \\   
&& \frac { (^3\text{He}/d)_{JAH} }  {(^3\text{He}/d)_{KI}} \approx \frac{N_p^{(DP)}}{N_p^{(KI)}} \approx \frac{1}{2}.   \label{eq:RHe3dCK}          
\end{eqnarray} }%
Eqs.~(\ref{eq:RdpCK}-\ref{eq:RHe3dCK}) explicitly show effects of the hadronic system evolution on light nucleus production. 
In order to better probe effects of hadronic evolution stage on the production of light nuclei, we construct the following observables at the final kinetic freeze-out,
i.e., values of these yield ratios in the case of shuting off hadronic interactions after the hadronization so that there are no light nucleus formation
during the hadronic system evolution to values of these ratios with hadronic interactions
{\setlength\arraycolsep{0.2pt}
\begin{eqnarray}
&& \frac { (d/p)_{noHI} }  {(d/p)_{withHI}} \approx \frac{1}{2}\frac { (d/p)_{JAH} }  {(d/p)_{KI}} \approx \frac{1}{4},   \label{eq:RdpHR} \\
&& \frac { (^3\text{He}/p)_{noHI} }   {(^3\text{He}/p)_{withHI}}  \approx  \frac{1}{2}\frac { (^3\text{He}/p)_{JAH} }   {(^3\text{He}/p)_{KI}}  \approx \frac{1}{8},  \label{eq:RHe3pHR} \\   
&& \frac { (^3\text{He}/d)_{noHI} }  {(^3\text{He}/d)_{withHI}} \approx \frac { (^3\text{He}/d)_{JAH} }  {(^3\text{He}/d)_{KI}} \approx \frac{1}{2}.   \label{eq:RHe3dHR}          
\end{eqnarray} }%
These results show that values for $d/p$, $^3$He$/p$ and $^3$He$/d$ when shuting off hadronic interactions decrease to about 1/4, 1/8 and 1/2 compared to those with hadronic interactions.
This decrease effect has been observed in pp collisions at $\sqrt{s}=7$ TeV by $d/p$, which is lower by a factor of 2.2 than in Pb-Pb collisions \cite{deuPRC2016}.
Further measurements for $^3$He$/p$ and $^3$He$/d$ in pp collisions can help to confirm this decrease effect of the hadronic interactions in the hadronic system evolution.

\begin{figure}[htbp]
\centering
 \includegraphics[width=1.\linewidth]{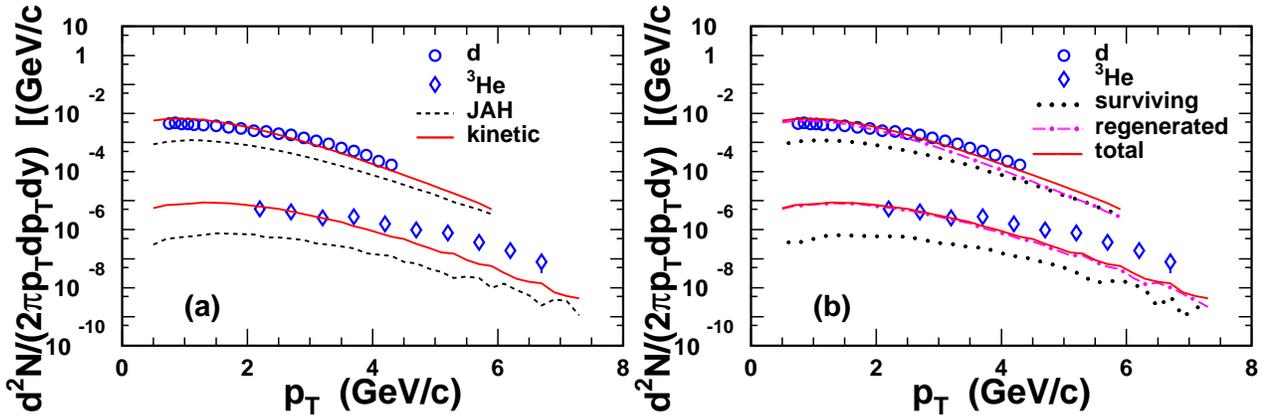}\\
 \caption{(Color online) $p_T$ distributions of $d$ and $^3$He at midrapidity in central Pb-Pb collisions at $\sqrt{s_{NN}}=2.76$ TeV.
 The open cycles and diamonds are the experimental data for $d$ and $^3$He from Ref. \cite{deuPRC2016}.
 (a) The solid lines are the results at the final kinetic freeze-out and the dashed lines are for those just after the hadronization.
 (b) The solid, dotted and dash-dotted lines are the results for the total, surviving and regenerated light nuclei at the final kinetic freeze-out, respectively.}
 \label{centraldeuteronPT}
\end{figure}

$p_T$ distributions of $d$ and $^3$He are studied and the results are in Fig.~\ref{centraldeuteronPT}.
The open cycles and diamonds are the experimental data for $d$ and $^3$He, respectively, measured by the ALICE Collaboration in Ref. \cite{deuPRC2016}.
In Fig.~\ref{centraldeuteronPT} (a), the results calculated just after the hadronization are exhibited by the dashed lines and those at the final kinetic freeze-out by the solid lines.
For both $d$ and $^3$He, $p_T$ spectra calculated just after the hadronization are harder than those at the final kinetic freeze-out.
This is because the $p_T$ distribution of protons from $\Delta$ decay contribution is softer than that of the directly produced protons.
In Fig.~\ref{centraldeuteronPT} (b), $p_T$ distributions of the surviving $d$ and $^3$He combined by the directly produced nucleons and 
those regenerated by the nucleons decayed from $\Delta$ resonances or by combining directly produced nucleons with those from $\Delta$ resonance decays at the final kinetic freeze-out 
are presented with the dotted lines and the dot-dashed lines, respectively. 
The solid lines denoting the results for the total $d$ and $^3$He at the kinetic freeze-out are still printed for guide.

\section{summary}

We have proposed a new method, an exclusive quark combination model followed by an inclusive hadron recombination model, 
to systematically study different production sources of light nuclei in relativistic heavy-ion collisions.
Using the exclusive quark combination model SDQCM, we can obtain momentum distributions of identified hadrons for directly produced ones and those at the final kinetic freeze-out
which are necessary inputs for the following inclusive hadron recombination model.
In the hadron recombination model, we analytically deduced the formation formulas for $d$ and $^3$He and explicitly presented the used assumptions 
such as the coordinate and the momentum factorization for kernel functions and nucleon joint distributions, and the decoupling of different coordinate variables in the coordinate distributions.
We computed $dN/dy$ and $p_T$ spectra of $d$ and $^3$He as well as their ratios just after the hadronization and at the final kinetic freeze-out in Pb-Pb collisions at $\sqrt{s_{NN}}=2.76$ TeV.
Light nuclei produced by the directly produced nucleon combination and those combined by the nucleons from $\Delta$ resonance decays (or those by combining nucleons from decays with directly produced ones)
were distinguished in our method.

The calculated results showed that only very small fractions of light nuclei can be created just after the hadronization, and most of light nuclei were regenerated at the late stage of the hadronic system evolution.
Nucleons from $\Delta$ resonance decays contribute much to the regeneration of light nuclei, and these regenerated proportions were about 79\% and 92\% for $d$ and $^3$He, respectively, at the final kinetic freeze-out.
Ratios such as $d/p$, $^3$He$/p$ and $^3$He$/d$ were found to be effective observables to probe different production sources of light nuclei.
Values of $d/p$ and $^3$He$/d$ just after the hadronization, i.e., only including contributions from directly produced nucleons are about a half of those at the final kinetic freeze-out, i.e., 
including contributions both from nucleons coming from $\Delta$ decays and from directly produced ones, and for $^3$He$/p$ about 1/4.
$d/p$, $^3$He$/p$ and $^3$He$/d$ were also found to be effective probes for effects of hadronic interactions on light nucleus formation after the hadronization.
Values of $d/p$ and $^3$He$/p$ when shuting off hadronic interactions are about 1/4 and 1/8 of those considering hadronic interactions, and for $^3$He$/d$ about 1/2.
All of these results set new insights into the understanding for the production of light nuclei in relativistic heavy-ion collisions.

\section*{Acknowledgements}

The authors thank Qiang Zhao, Lie-Wen Chen and Kai-Jia Sun for helpful discussions.
This work is supported in part by the National Natural Science Foundation of China under Grant Nos. 11505104, 11575100 and 11675091.


\begin{thebibliography}{00}            

\bibitem{Gosset1977PRC}
J. Gosset, H. H. Gutbrod, G. Meyer, A. M. Poskanzer, A. Sandoval, R. Stock, and G. D. Westfa11, Phys.\ Rev.\ C {\bf16}, 629 (1977).
\bibitem{Mekjian1978PRC}
A. Z. Mekjian, Phys.\ Rev.\ C {\bf17}, 1051 (1978).
\bibitem{Kapusta1980PRC}
Joseph I. Kapusta, Phys.\ Rev.\ C {\bf21}, 1301 (1980).
\bibitem{Nagle1996PRC}
J. L. Nagle, B. S. Kumar, D. Kusnezov, H. Sorge, and R. Mattiello, Phys.\ Rev.\ C {\bf53}, 367 (1996).
\bibitem{LWChen2003NPA}
L. W. Chen, C.M. Ko, and B. A. Li, Nucl.\ Phys.\ A {\bf729}, 809 (2003).
\bibitem{STARPRL2001}
C. Adler {\it et al.} (STAR Collaboration), Phys.\ Rev.\ Lett. {\bf87}, 262301 (2001).
\bibitem{PHENIXPRL2007}
S. Afanasiev {\it et al.} (PHENIX Collaboration), Phys.\ Rev.\ Lett. {\bf09}, 052301 (2007).
\bibitem{STARarXiv2009}
B. I. Abelev {\it et al.} (STAR Collaboration), arXiv:0909.0566 [nucl-ex].
\bibitem{STARPRC2016}
L. Adamczyk {\it et al.} (STAR Collaboration), Phys.\ Rev.\ C {\bf94}, 034908 (2016).
\bibitem{STARarXiv2017}
Ning Yu (for the STAR Collaboration), arXiv:1704.04335 [nucl-ex].
\bibitem{Bearden2002EPJC}
I. G. Bearden {\it et al.}, Eur.\ Phys.\ J.\ C {\bf23}, 237 (2002).
\bibitem{NA49PLB2000}
S. V. Afanasev {\it et al.} (NA49 Collaboration), Phys.\ Lett. B {\bf486}, 22 (2000).
\bibitem{NA49PPC2004}
T. Anticic {\it et al.} (NA49 Collaboration), Phys.\ Rev.\ C {\bf69}, 024902 (2004).
\bibitem{NA49PPC2012}
T. Anticic {\it et al.} (NA49 Collaboration), Phys.\ Rev.\ C {\bf85}, 044913 (2012).
\bibitem{NA49PPC2016}
T. Anticic {\it et al.} (NA49 Collaboration), Phys.\ Rev.\ C {\bf94}, 044906 (2016).
\bibitem{deuPRC2016}
J. Adam {\it et al.} (ALICE Collaboration), Phys.\ Rev.\ C {\bf93}, 024917 (2016).
\bibitem{JPCS2015}
J. Anielski (for the ALICE Collaboration), J. Phys.\ Conf.\ Ser.\ {\bf612}, 012014 (2015).
\bibitem{ALICE2016NPA}
Ramona Lea (for the ALICE Collaboration), Nucl.\ Phys.\ A {\bf956}, 264 (2016).
\bibitem{JPCS2017}
Stefano Trogolo (for the ALICE Collaboration), J. Phys.\ Conf.\ Ser.\ {\bf832}, 012061 (2017).
\bibitem{FLShao2017PRC}
F. L. Shao, G. J. Wang, R. Q. Wang, H. H. Li and J. Song, Phys.\ Rev.\ C {\bf95}, 064911 (2017).
\bibitem{Csernai1986PR}
L. P. Csernai and J. I. Kapusta, Phys.\ Rep. {\bf131}, 223 (1986).
\bibitem{AndronicPLB2011}
A. Andronic, P. Braun-Munzinger, J. Stachel, and H. Stoecker, Phys.\ Lett. B {\bf697}, 203 (2011).
\bibitem{Chatterjee2014PRC}
S. Chatterjee and B. Mohanty, Phys.\ Rev.\ C {\bf90}, 034908 (2014).
\bibitem{Butler1963PR}
S. T. Butler and C.A. Pearson, Phys.\ Rev. {\bf129}, 836 (1963).
\bibitem{Schwarzschild1963PR}
A. Schwarzschild and C. Zupancic, Phys.\ Rev. {\bf129}, 854 (1963).
\bibitem{SatoPLB1981}
H. Sato and K. Yazaki, Phys.\ Lett. B {\bf98}, 153 (1981).
\bibitem{Dover1991PRC}
C. B. Dover, U. Heinz, and E. Schnedermann, Phys.\ Rev.\ C {\bf44}, 1636 (1991).
\bibitem{YGMa2012PRC}
L. Xue, Y. G. Ma, J. H. Chen, and S. Zhang, Phys.\ Rev.\ C {\bf85}, 064912 (2012).
\bibitem{GChen2013PRC}
G. Chen, H. Chen, J. Wu, D. S. Li, and M. J. Wang, Phys.\ Rev.\ C {\bf88}, 034908 (2013).
\bibitem{LWChenPLB2015}
K. J. Sun and L. W. Chen, Phys.\ Lett. B {\bf751}, 272 (2015).
\bibitem{LLZhu2017PRC}
X. J. Yin, C. M. Ko, Y. F. Sun, and L. L. Zhu, Phys.\ Rev.\ C {\bf95}, 054913 (2017).
\bibitem{JPCS2016}
S. Bufalino (for the ALICE Collaboration),  J. Phys.\ Conf.\ Ser.\ {\bf718}, 042008 (2016).
\bibitem{QBXie1988PRD}
Q. B. Xie and X. M. Liu , Phys.\ Rev.\ D {\bf 38}, 2169  (1988).
\bibitem{FLShao2005PRC}
F. L. Shao, Q. B. Xie, and Q. Wang, Phys.\ Rev.\ C {\bf71}, 044903 (2005).
\bibitem{CEShao2009PRC}
C. E. Shao, J. Song, F. L. Shao, and Q. B. Xie,  Phys.\ Rev.\ C {\bf80}, 014909 (2009).
\bibitem{radii2009PRC}
G. Ropke, Phys.\ Rev.\ C {\bf79}, 014002 (2009).
\bibitem{Rche2017acta}
S. Mrowczynski, Acta Phys.\ Polon.\ B {\bf48}, 707 (2017).
\bibitem{Rkin2016PRC}
J. Adam {\it et al.} (ALICE Collaboration), Phys.\ Phys.\ C {\bf93}, 024905 (2016).
\bibitem{QWang2000PRD}
Q. Wang, G. Gustafson, and Q. B. Xie , Phys.\ Rev.\ D {\bf62}, 054004 (2000).
\bibitem{RQWang2012PRC}
R. Q. Wang, F. L. Shao, J. Song, Q. B. Xie, and Z. T. Liang, Phys.\ Rev.\ C {\bf86}, 054906 (2012).
\bibitem{JSong2017PRC}
J. Song, H. H. Li, R. Q. Wang, and F. L. Shao, Phys.\ Rev.\ C {\bf95}, 014901 (2017).
\bibitem{RQWang2015PRC}
R. Q. Wang, J. Song, and F. L Shao, Phys.\ Rev.\ C {\bf91}, 014909 (2015).
\bibitem{pikpPRL2012}
B. Abelev {\it et al.} (ALICE Collaboration), Phys.\ Rev.\ Lett. {\bf109}, 252301 (2012).
\bibitem{Ks0lam2013}
B. Abelev {\it et al.} (ALICE Collaboration), Phys. Rev. Lett. {\bf111}, 222301 (2013).
\bibitem{xiome2013}
B. Abelev {\it et al.} (ALICE Collaboration), Phys.\ Lett. B {\bf728}, 216 (2014).
\bibitem{phipt}
B. Abelev {\it et al.} (ALICE Collaboration), arXiv:1404.0495v2 [nucl-ex].
\bibitem{PDG2016}
C. Patrignani {\it et al.} (Particle Data Group), Chin.\ Phys.\ C {\bf40}, 100001 (2016).


\end{thebibliography}
\end{document}